\documentclass[twocolumn]{aastex631}

\shorttitle{}
\shortauthors{Arora \& Hasegawa}
\graphicspath{{./}{figures/}}
\usepackage{xcolor}
\usepackage{soul}
\begin{document}

\title{The architecture of multi-planet systems as a tracer of their formation mechanisms}

\author[0000-0001-7250-0862]{Udit Arora}
\affiliation{Indraprastha Institute of Information Technology, Okhla Industrial Estate, Phase III, New Delhi, 110020, India}

\author{Yasuhiro Hasegawa}
\affiliation{Jet Propulsion Laboratory, California Institute of Technology, Pasadena, CA 91109, USA}

\email{UA:udit18417@iiitd.ac.in; YH:yasuhiro.hasegawa@jpl.nasa.gov}



\begin{abstract}

Exoplanets observed by the {\it Kepler} telescope exhibit a bi-modal, radius distribution, which is known as the radius gap.
We explore an origin of the radius gap, focusing on multi-planet systems.
Our simple theoretical argument predicts that type I planetary migration produces different configurations of protoplanets with different masses
and such different configurations can result in two distinguishable populations of small-sized multi-planet systems.
We then perform an observational analysis to verify this prediction.
In the analysis, multiple Kolmogorov–Smirnov tests are applied to the observed systems,
using the statistical measures that are devised to systematically characterize the properties of multi-planet systems.
We find with 99.5\% confidence that the observed, small-sized multi-planet systems are divided into two distinct populations.
The distinction likely originates from different spatial distributions of protoplanets, which are determined by type I migration and subsequently trigger giant impact.
We also show that these distinct populations are separated around the radius gap 
when the gas surface density of protoplanetary disks is $\sim 10^2$ g cm$^{-2}$ in the vicinity of the host stars.
This work therefore emphasizes the importance of planetary migration and the inner disk properties.

\end{abstract}



\section{Introduction} \label{sec:intro}

The rapid increase of observed exoplanets revolutionizes our understanding of planet formation \citep[e.g.,][for a review]{2015ARA&A..53..409W},
which has been accelerated thanks to the {\it Kepler} mission \citep{2010Sci...327..977B}.
One famous, astonishing result is that small-sized planets whose radius is larger than $1 R_{\oplus}$ and smaller than $ 4R_{\oplus}$, 
are dominant in the observed population \citep[e.g.,][]{2010Sci...330..653H}.
This is consistent with the pioneering discovery of radial velocity observations \citep{2011arXiv1109.2497M}.

The ubiquity of observed small-sized planets has challenged the canonical picture of planet formation \citep[e.g.,][]{2004ApJ...604..388I}.
This is because there is no such analogue in the solar system,
and hence both the observational characterization of these planets and 
the theoretical development of possible formation mechanisms have been required.
For the former, the follow-up observations of exoplanet-host stars reveal that 
the radius range of small-sized planets is divided into two regimes \citep{2017AJ....154..109F}: 
$1 R_{\oplus} \la R_p \la 1.8  R_{\oplus}$ and $1.8 R_{\oplus} \la R_p \la 4  R_{\oplus}$. 
Based on the bulk density \citep[e.g.,][]{2014ApJ...783L...6W,2015ApJ...801...41R}, 
planets in the first regime can literally be called as ``super-Earth", 
and those in the second regime may be named as ``sub-Neptune". 
The presence of these bi-modal populations has been popularized as the "radius gap (or valley)", and triggered a number of theoretical studies.
Recent studies suggest that both the formation and evolution processes can generate the gap \citep[e.g.,][]{2018MNRAS.476..759G,2013ApJ...775..105O}.

The multiplicity is another important feature of observed small-sized planets \citep[e.g.,][]{2011Natur.470...53L},
which also sheds light on their formation mechanisms \citep[e.g.,][]{2013ApJ...775...53H,2016ApJ...822...54D}. 
Intriguingly, \citet{2018AJ....155...48W} confirm the presence of the radius gap, focusing only on exoplanets in multi-planet systems.
This finding may suggest that physical processes inevitable for shaping multi-planet systems would contribute to the generation of the radius gap.

Here, we show through a simple theoretical argument and an observational analysis that 
planetary migration produces different configurations of protoplanets with different masses
and that this difference and the subsequent giant impact lead to two distinguishable populations of small-sized multi-planet systems.
It is prominent that the dividing radius of these two populations broadly corresponds to the radius gap
when the gas surface density of the natal protoplanetary disk has a low value ($\sim 10^2$ g cm$^{-2}$) in the vicinity of the central star.
Thus, this work points out the importance of planetary migration and the inner disk properties
to better understand the observed properties of super-Earths and sub-Neptunes.

\section{Theoretical and Observational Analyses} \label{sec:mod}

\subsection{Exoplanet data} \label{sec:data}

We first introduce the exoplanetary data that will be used in the following analyses; 
in these analyses, the mass ($M_p$), radius ($R_p$), semimajor axis ($a_p$) of planets and the mass ($M_s$) of the host stars are needed.

We obtain the {\it Kepler} data from \citet{keplercumulativetable}\footnote{
We accessed to https://exoplanetarchive.ipac.caltech.edu/cgi-bin/TblView/nph-tblView?app=ExoTbls\&config=cumulative
on 2021-01-21 at 16:17, and obtained the data with the size of 9564 $\times$ 49 columns.} 
We focus only on multi-planetary systems orbiting around single stars,
which are comprised of both confirmed and candidate planets.
We filter out the systems that do not have certain quantities and hence our analyses cannot be applied to.
These are the surface gravity, effective temperature, and radius of the host stars, and the radius and orbital period of planets.
Note that the stellar parameters are needed both to identify the host stars as the main sequence ones, and to compute their masses;
for the former, we follow the approach of \citet{2018AJ....156...24M},
where dwarf stars are identified, based on the relationship between the surface gravity and effective temperature \citep{2016ApJS..224....2H}.
For the latter, we use the mass-luminosity and the mass-radius relations \citep{2018MNRAS.479.5491E}.\footnote{
Note that the stellar masses estimated by our approach are comparable to those obtained, using the Gaia data \citep{2020AJ....159..280B};
we find that the difference is small ($< 20$ \%) enough for most cases that our results do not change very much.}

We select planets whose radius is in the range of $0.1 R_{\oplus} \leq R_p  \leq 30 R_{\oplus}$.
Following \citet{2018ApJS..235...38T}, we adopt the disposition score of $> 0.6$,
where the trade-off between reliability and completeness is achieved properly (see the top panel of their figure 13).
We have confirmed that our results do not change very much even if planets with the disposition score of $> 0.5$ are chosen.
The semimajor axis of planets is computed, using the Kepler's third law. 

In summary, we have the total of 348 planetary systems with 870 exoplanets observed by {\it Kepler}.\footnote{
Note that we have removed one planetary system, Kepler-1659, due to the unphysically high mass of one planet.}

\subsection{Mass-radius relation}

The planet mass is one fundamental quantity for our analyses and is computed from the mass-radius relation.

We utilize the pretrained model developed by \citet{2017ApJ...834...17C},
where an unbiased forecasting model was built upon for a probabilistic mass-radius relation with the Bayesian framework.
They found that the mass-radius relation is divided into two regimes for small-sized planets:
$ M_p \la 2  M_{\oplus}$, and $2 M_{\oplus} \la M_p \la 100  M_{\oplus}$,
and the corresponding regimes are referred to as the Terran worlds, and the Neptunian worlds, respectively.

\begin{figure}
\begin{center}
\includegraphics[width=8cm]{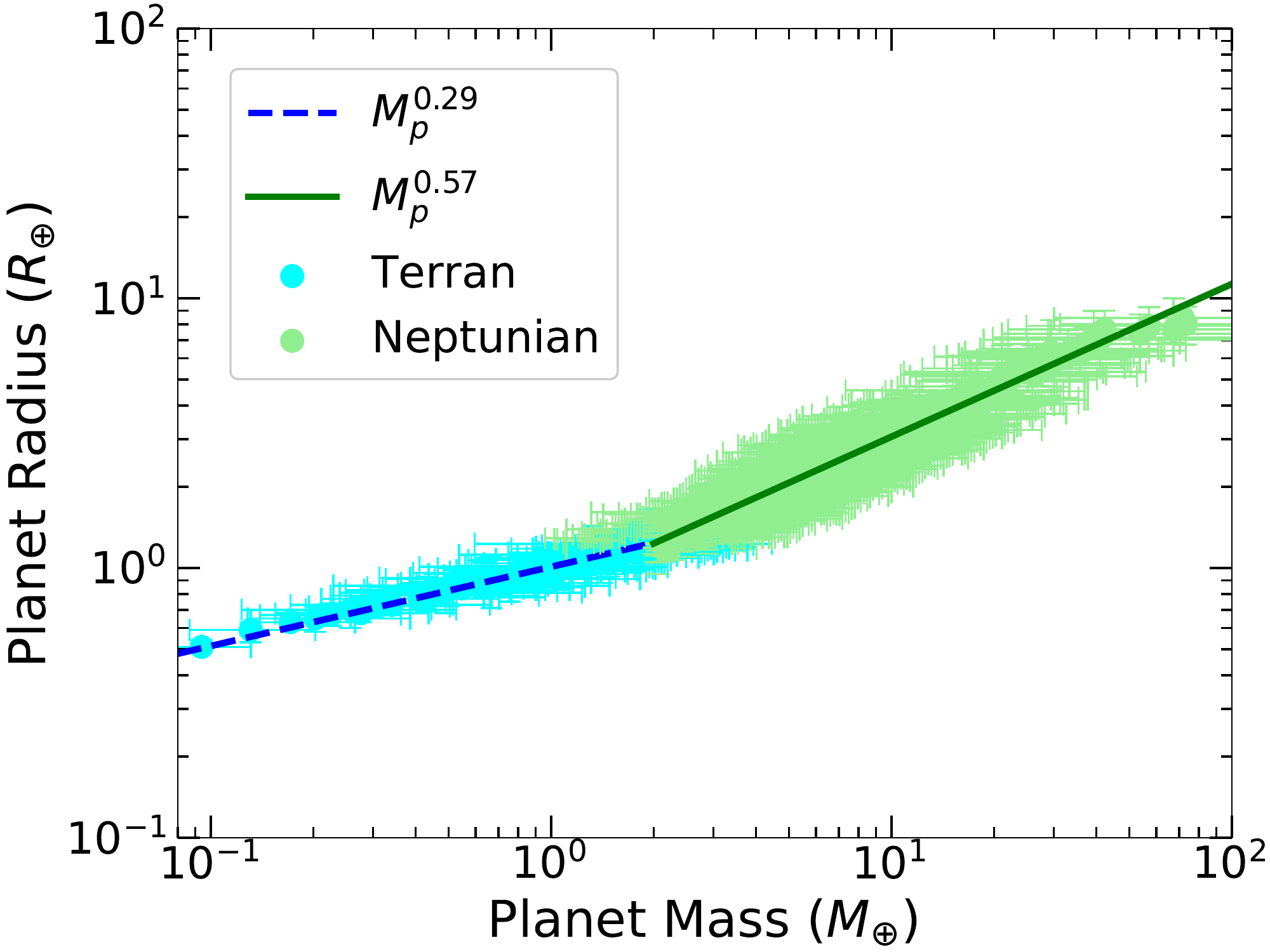}
\caption{The mass-radius relation for exoplanets in our data set. 
The cyan dots with the error bars represent planets in the Terran world (i.e., $M \la 2M_{\oplus}$), 
and the light green dots with the error bars represent planets in the Neptunian worlds (i.e., $2M_{\oplus} \la  M \la100 M_{\oplus}$).
The corresponding fitting profiles are denoted by the blue dashed, and the green solid lines, respectively.}
\label{fig1}
\end{center}
\end{figure}

Figure \ref{fig1} shows the computed masses for given radii.
We subsequently fit these data and confirm that
the resulting power-law indices for the two types of planets are similar to those of \citet{2017ApJ...834...17C}.
It is interesting that the Neptunian worlds reside in the radius range of $1.2 R_{\oplus} \la R_p \la 10  R_{\oplus}$,
and therefore they cover both (massive) super-Earths and the entire sub-Neptunes.

\subsection{Theoretical Prediction: the importance of planetary migration} \label{sec:theory}

The formation of small-sized, multi-planet systems is currently the target of active research 
\citep[e.g., ][]{2010ApJ...719..810I,2013ApJ...775...53H,2013MNRAS.431.3444C,2017MNRAS.470.1750I}, 
and it is still far from a complete understanding.

Our theoretical argument assumes that type I migration and the subsequent giant impact play an important role in forming small-sized, multi-planet systems;
type I migration takes place due to the disk-planet interaction and becomes effective for low-mass planets \citep[$M_p \ga 1M_{\oplus}$, e.g.,][]{2012ARA&A..50..211K}.
Giant impact is the collision between protoplanets and serves as the final mass assembly for terrestrial planets in the solar system \citep[e.g.,][hereafter C01]{2001Icar..152..205C}.

Under this assumption, the importance of planetary migration to trigger the subsequent giant impact between neighbouring protoplanets may be determined 
by the difference between the mutual spacing ($\Delta_{\rm mig}$) achieved by type I migration and that ($\Delta_{\rm esc}$) by the escape velocity;
$\Delta_{\rm esc}$ is the result of the pure gravitational interaction between the neighbouring protoplanets 
and is the largest separation between them before they undergo giant impact \citep[e.g.,][]{2014ApJ...795L..15S}:
\begin{equation}
\Delta_{\rm esc} \simeq 2 a_p \frac{v_{\rm esc}}{v_{\rm Kep}},
\end{equation}
where $v_{\rm Kep}$ is the Keplerian velocity, and $v_{\rm esc}= \sqrt{2GM_p/R_p}$ is the escape velocity for the neighbouring protoplanets with comparable masses and radii.

Planetary migration affects the onset of the subsequent giant impact considerably if the following condition is met:
\begin{equation}
\Delta_{\rm mig} < \Delta_{\rm esc}.
\end{equation}
Mathematically, $\Delta_{\rm mig}$ can be computed from the consideration that 
the speed ($v_{\rm mig}$) of (differential) type I migration, which reduces the mutual spacing ($b=|a_{p,i}-a_{p,j}|$) between the neighbouring protoplanets ($i,j$),
is compensated by the gravitational repulsion between them \citep{2010ApJ...719..810I}.
For $v_{\rm mig}$, it is written as \citep[e.g., ][]{2011MNRAS.417.1236H}
\begin{eqnarray}
v_{\rm mig}   \simeq 2 K_{\rm mig} \frac{\Sigma_{\rm g} a_p^2 M_p}{M_s^2} \left( \frac{H}{a_p} \right)^{-2} v_{\rm Kep},
\end{eqnarray}
where $\Sigma_{\rm g}$ and $H$ are the gas surface density and the pressure scale height of the natal protoplanetary disk, respectively, 
and $K_{\rm mig}$ is a coefficient which accounts for the differential speed and the detail of disk properties.
For simplicity, we here adopt that $H/a_{p}=0.02$, $K_{\rm mig}=1$, and $M_{s} = 1 M_{\odot}$.
For the gravitational repulsion, the expansion ($\delta b$) of the mutual spacing is given by linear theory as \citep{1990A&A...227..619H}
\begin{equation}
\delta b \simeq 30 \left( \frac{b}{r_{\rm H}} \right)^{-5} r_{\rm H},
\end{equation}
where $r_{\rm H} = a (2M_{p}/3M_s)^{1/3}$ is the mutual Hill radius for the neighbouring protoplanets with comparable masses.
Since the expansion is driven by encounters which occur at every synodic period ($T_{\rm syn}$),
the expansion rate becomes 
\begin{equation}
 \frac{d b}{dt} \simeq \frac{\delta b}{T_{\rm syn}},
\end{equation}
where $T_{\rm syn} \simeq 4 \pi a_p^2 / 3b v_{\rm Kep}$.
In summary, $\Delta_{\rm mig} $ can be computed, by equating $v_{\rm mig}$ with $\delta b/T_{\rm syn}$.

\begin{figure}
\begin{center}
\includegraphics[width=8cm]{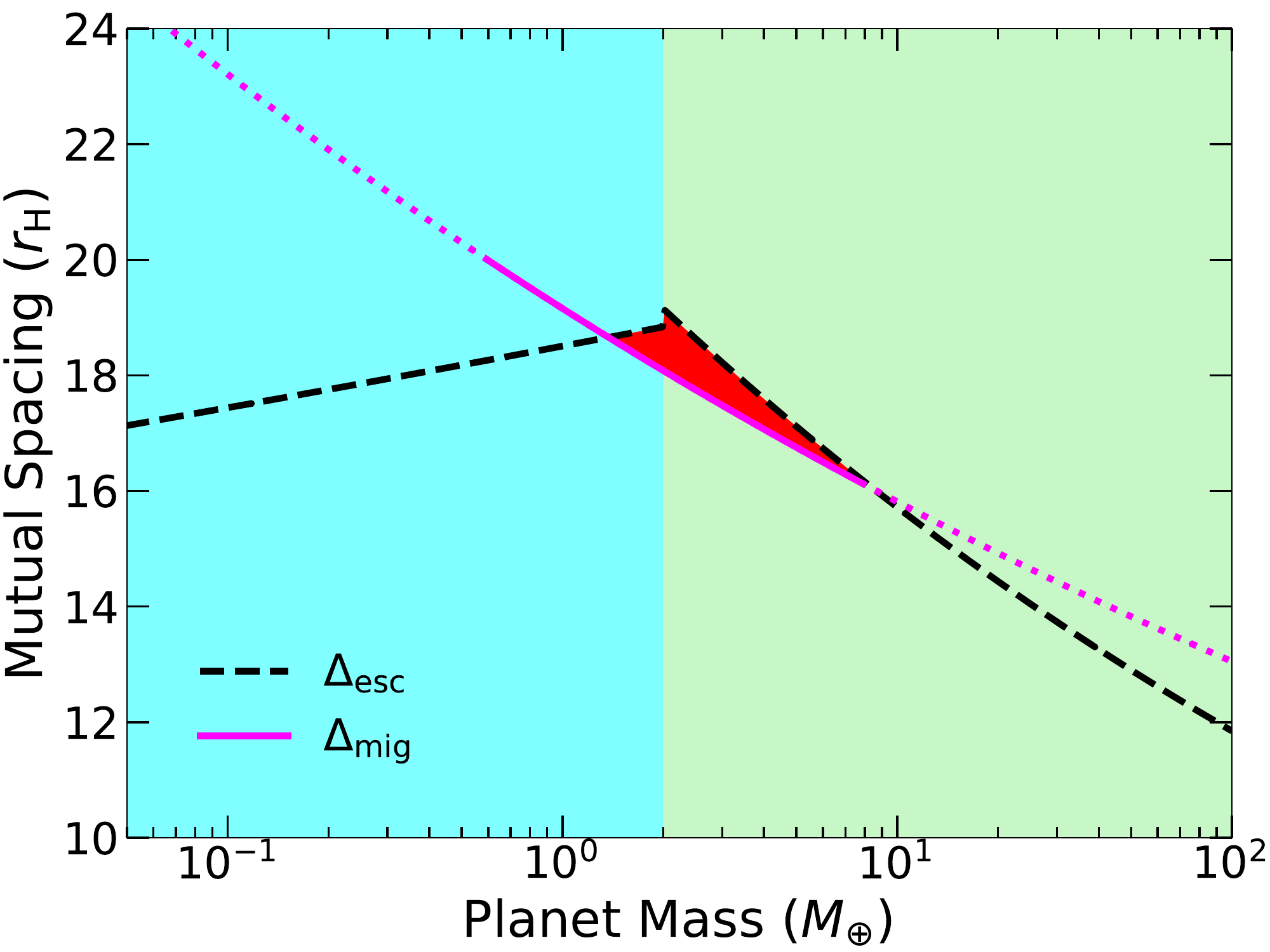}
\caption{The mutual spacing as a function of planet mass for the case that $\Sigma_{\rm g}= 5 \times 10^{2}$ g cm$^{-2}$ at $a_p=0.1$ au. 
Planetary migration achieves narrower mutual spacings for the mass range of $1 M_{\oplus} \la M_p \la 10 M_{\oplus}$ (see the red shaded region).
The mass-radius relation is used to plot $\Delta_{\rm esc}$ (see the black, dashed line).
For $\Delta_{\rm mig}$, the solid line denotes  the mass regime where type I migration becomes effective (see the magenta line).
The background cyan and light green regions represent the Terran and Neptunian worlds, respectively.}
\label{fig2}
\end{center}
\end{figure}

Figure \ref{fig2} shows the resulting behaviors of $\Delta_{\rm esc}$ and $\Delta_{\rm mig}$ as a function of planet mass.
In this plot, the mass-radius relation obtained in Figure \ref{fig1} is used to compute $\Delta_{\rm esc}$.
As an example, we adopt the value of $\Sigma_{\rm g}= 5 \times 10^{2}$ g cm$^{-2}$ at $a_p=0.1$ au to calculate $\Delta_{\rm mig}$.
In addition, the mass regime where type I migration becomes effective is denoted by the solid line of $\Delta_{\rm mig}$,
following \citet{2012ApJ...760..117H}.

We find that $\Delta_{\rm mig} > \Delta_{\rm esc}$ for the planet mass of  $M_p \la 1 M_{\oplus}$,
 $\Delta_{\rm mig} < \Delta_{\rm esc}$ for $1 M_{\oplus} \la M_p \la 10 M_{\oplus}$,
and $\Delta_{\rm mig} \simeq \Delta_{\rm esc}$ for $M_p \ga 10 M_{\oplus}$ in this particular setup.
Note that the mutual spacing of $20 r_{\rm H}$ corresponds to the period ratio of $\sim 1.5$,
which is consistent with the peak value of the observed exoplanets \citep[e.g.,][]{2020MNRAS.495.4192C}.
Our value is, however, larger than what the typical numerical simulations predict \citep[i.e., $\sim$ a few $r_{\rm H}$, e.g.,][]{2017MNRAS.470.1750I}.
This difference comes from a lower value of $\Sigma_{\rm g}$ (see Section \ref{sec:disc} for more discussion).

Given that the mutual spacing is one key parameter for giant impact,
it is natural to consider that 
protoplanets with $\Delta_{\rm mig} \la \Delta_{\rm esc}$ undergo giant impact more efficiently/rapidly than those in $\Delta_{\rm mig} > \Delta_{\rm esc}$.
This consideration leads to the expectation that (proto)planets in the red shaded region may surely experience giant impact and obtain higher masses.
Note that the size of the region depends on the power-law index of the mass-radius relation;
assuming that $R_p \propto M_p^{\beta}$, $\Delta_{\rm mig} \propto M^{-1/12}$ and $\Delta_{\rm esc} \propto M^{(1/3-\beta)/2}$ in the unit of $r_{\rm H}$. 
The region expands if $\beta > 1/3$ and shrinks if  $\beta < 1/3$ in the Neptunian worlds, compared with Figure \ref{fig2}.
However, the region surely exists if $\Sigma_{\rm g}$ is relatively small (see Section \ref{sec:disc}).

In the following, we conduct the observational analysis to verify our theoretical prediction.

\subsection{Statistical measures}

The direct observabales (e.g., the period ratio of neighboring planets) can be used to characterize multi-planet systems \citep[e.g.,][]{2018AJ....155...48W,2018ApJ...860..101Z}.
However, as the number of planets in the systems increases, the complexity in systematically characterizing multi-planet systems as a whole increase \citep[e.g.,][]{2020AJ....159..281G}. 

In order to resolve this issue, we adopt the so-called "statistical measures" that are devised by C01.
We here focus on two quantities that are computed directly from $M_p$, $a_p$, and $M_s$.

The first quantity is the mass concentration ($S_c$) that measures the degree to which the planet mass is concentrated in a certain location of the system
and is calculated as
\begin{equation}
    S_c = \mbox{max} \left( \frac{\sum_j  M_{p,j}}{\sum_j M_{p,j} [ \log_{10}{(a/a_{p,j})]^2}} \right),
\label{mc}
\end{equation}
where the summation is done for the index $j$.
By changing $a$, the maximum value is searched.
Based on $N$-body simulations,
the value of $S_c$ reflects the initial spatial distribution of protoplanets that undergo giant impact eventually 
\citep[hereafter, H09 and HM12, respectively]{2009ApJ...703.1131H,2012ApJ...751..158H}.

The second quantity is the orbital spacing ($S_s$) which is somewhat similar to the averaged mutual spacing normalized by the mutual Hill radius.
The main difference is that $S_s$ is normalized by $M_p^{1/4}$ (not $M_p^{1/3}$) 
and is motivated by the results of $N$-body simulations that explore the stability of multi-planet systems \citep{1996Icar..119..261C}:
\begin{equation}
    S_s = \frac{6}{N-1} \Bigg( \frac{a_{p,max} - a_{p,min}}{a_{p,max}+a_{p,min}}\Bigg)\Bigg(\frac{3 M_s}{2\ \overline{M}_p}\Bigg)^{1/4},
\label{oss}    
\end{equation}
where $N$,  $a_{p,max}$, $a_{p,min}$, and $\overline{M}_p$ are the number, the maximum and the minimum values of the semi-major axis, 
and the mean mass of the planets in a system, respectively.

\subsection{Observational Analysis: KS tests}

\begin{figure*}
\begin{minipage}{17cm}
\begin{center}
\includegraphics[width=8.5cm]{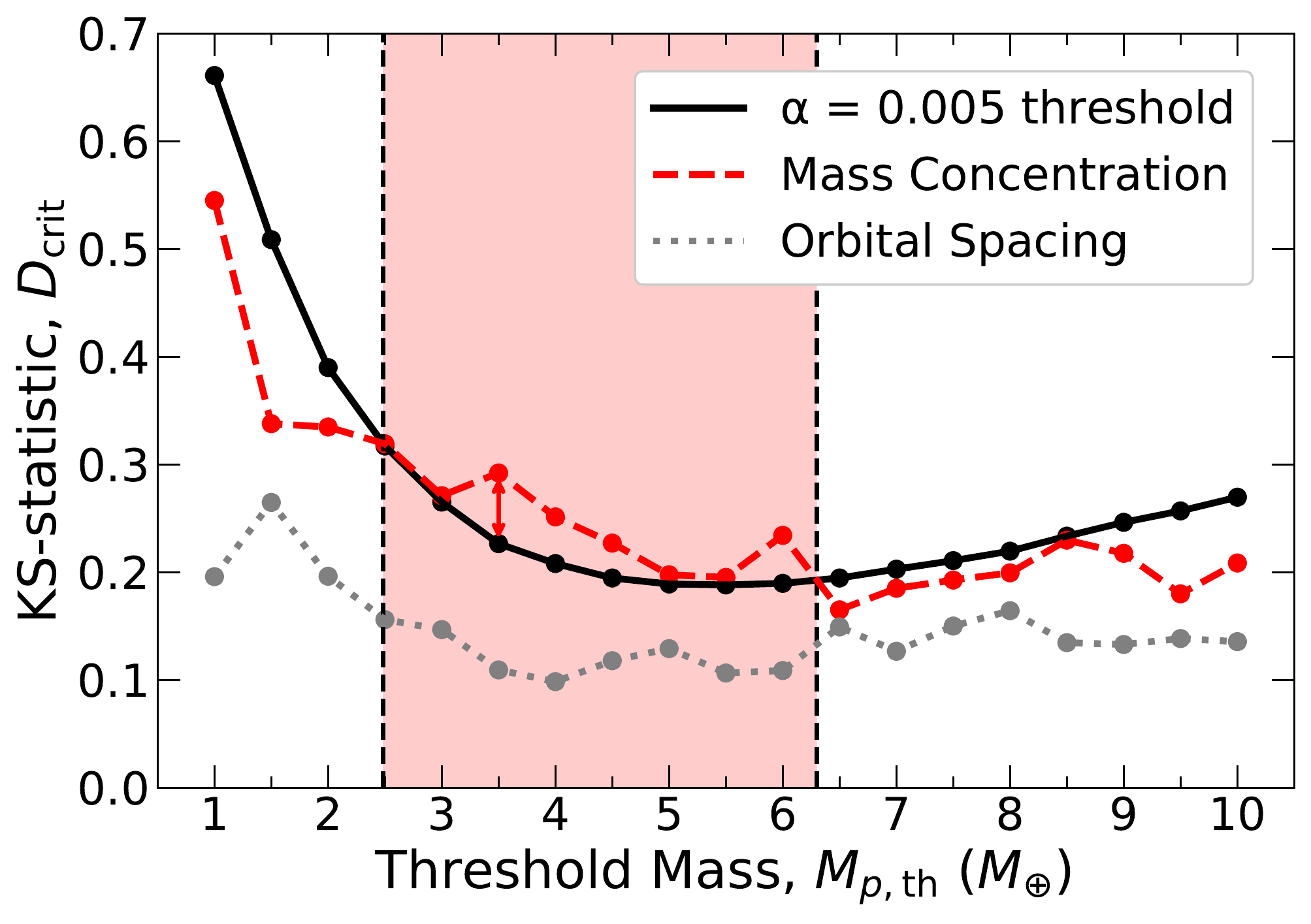}
\includegraphics[width=8cm]{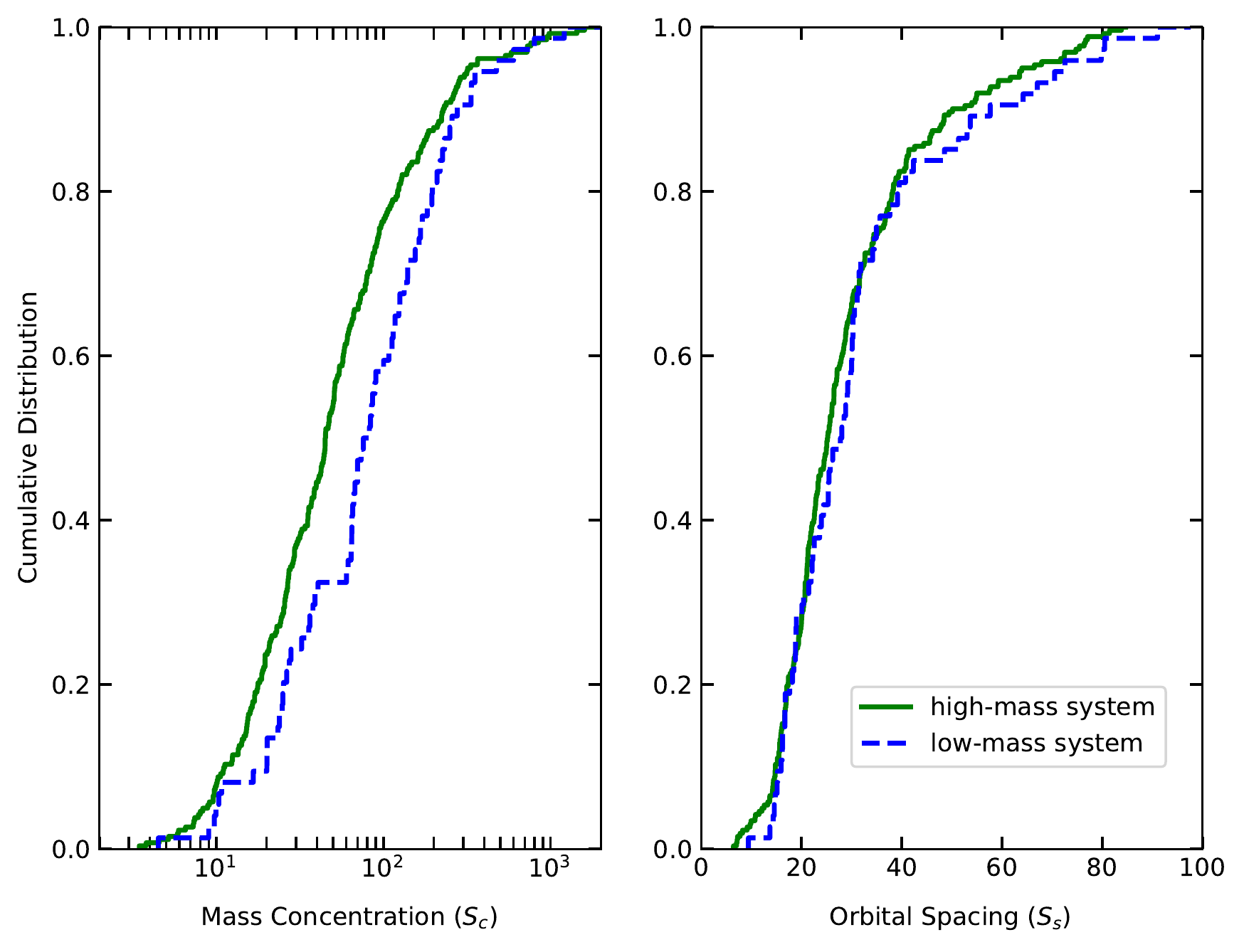}
\caption{The results of our KS tests and the best case of the cumulative distributions of $S_c$ and $S_s$ (i.e., $M_{p,\rm th}=3.5M_\oplus$).
The left panel shows that $S_c$ can be used to divide the full sample of small-sized, multi-planet systems into two populations 
in the mass range of $2.5 M_{\oplus} \la M_{p, \rm th} \la 6.3 M_{\oplus}$ (see the red shaded region).
In this region, the KS statistic (the red dashed line) is greater than the critical value (the black solid line),
and hence the null hypothesis is rejected with the 99.5\% confidence;
equivalently, low-mass ($M_{p,\rm ave} \leq M_{p, \rm th}$) and high-mass ($M_{p,\rm ave} > M_{p,\rm th}$) systems are not drawn from the same distribution.
On the contrary, $S_s$ does not show the similar feature.
The central and right panels show the best case,
where low-mass and high-mass systems are denoted by the blue dashed and the green solid lines, respectively.}
\label{fig3}
\end{center}
\end{minipage}
\end{figure*}

We now perform the Kolmogorov–Smirnov test (KS test) for the cumulative distributions of $S_c$ and $S_s$ 
to determine whether the population of small-sized, multi-planet systems can be divided into sub-groups.

The KS test is a non-parametric test and used to discover whether two samples are drawn from the same distribution, based on the null hypothesis.
The significance level (also referred to as $\alpha$) is the probability of rejecting the null hypothesis when it is actually true. 
The KS test for the two samples returns a KS statistic ($\mathcal{K}$) and a p-value ($\mathcal{P}$). 
Based on the value of $\alpha$, we calculate a critical value ($D_{\rm crit} $), 
which is given as
\begin{equation}
    D_{\rm crit} = c(\alpha) \cdot \sqrt{\Bigg(\frac{1}{n_1} + \frac{1}{n_2}\Bigg)},
\label{dcrit}
\end{equation}
where $c(\alpha) = \sqrt{- \ln(\alpha/2)/2}$, and $n_1$ and $n_2$ are the two sample sizes we employ. 
If $\mathcal{K}>D_{\rm crit}$ or $\mathcal{P}<\alpha$, then we can reject the null hypothesis at a level of $\alpha$.

We choose the average mass of planets in the systems ($M_{p, \rm ave}$) as a parameter to divide the full samples into two groups;
the current {\it Kepler} data suggest that at least the planet radius difference in the systems may be insignificant \citep{2018AJ....155...48W}.
This does not necessarily mean that the planet mass difference is small as well (see Figure \ref{fig1}).
However, the mass-radius relation is one best tool currently available in the literature to estimate the planet mass,
and hence we use the planet mass for our KS tests.
Practically, we bin the mass range of $1 M_{\oplus} \leq M_{p, \rm th} \leq 10 M_{\oplus}$ with the bin size of $0.5 M_{\oplus}$,
where $M_{p,\rm th}$ is a parameterized threshold mass to divide the full samples into two groups,
and compare these two samples: planetary systems with $M_{p,\rm ave} \leq M_{p, \rm th}$ and those with $M_{p,\rm ave} > M_{p,\rm th}$.

Figure \ref{fig3} summarizes our results;
the left panel shows the results of our KS tests, and the central and right ones depict the cumulative distributions of $S_c$ and $S_s$ for the best case (i.e., $M_{p, \rm th}=3.5 M_{\oplus}$).
We find that for the mass concentration,
the null hypothesis can be rejected for $2.48 M_{\oplus} < M_{p, \rm th} < 6.3 M_{\oplus}$ with the 99.5\% confidence (i.e., $\alpha = 0.005$);
this is evident from the KS statistic which is greater than the value of $D_{\rm crit}$ in this range.
Our analysis therefore suggests that the full samples can be divided into two groups at $M_{p, \rm th}=3.5 M_{\oplus}$.
On the other hand, similar distinction is not possible for the orbital spacing.

We discuss below how our observational analysis is consistent with our theoretical prediction, 
and why the mass concentration and orbital spacing behave differently.

\section{Discussion} \label{sec:disc}

Our theoretical argument has suggested
that planetary migration should have played a more important role for planets with $M_{p} \ga 2 M_{\oplus}$ (Figure \ref{fig2}),
and our observational analysis has shown 
that planetary systems with $M_{p, \rm ave} \la 3 M_{\oplus}$ have a higher value of $S_c$ than those with $M_{p, \rm ave} \ga 3 M_{\oplus}$ (Figure \ref{fig3}).
Based on these results, we here discuss how small-sized, multi-planet systems form.

\begin{figure}
\begin{center}
\includegraphics[width=8cm]{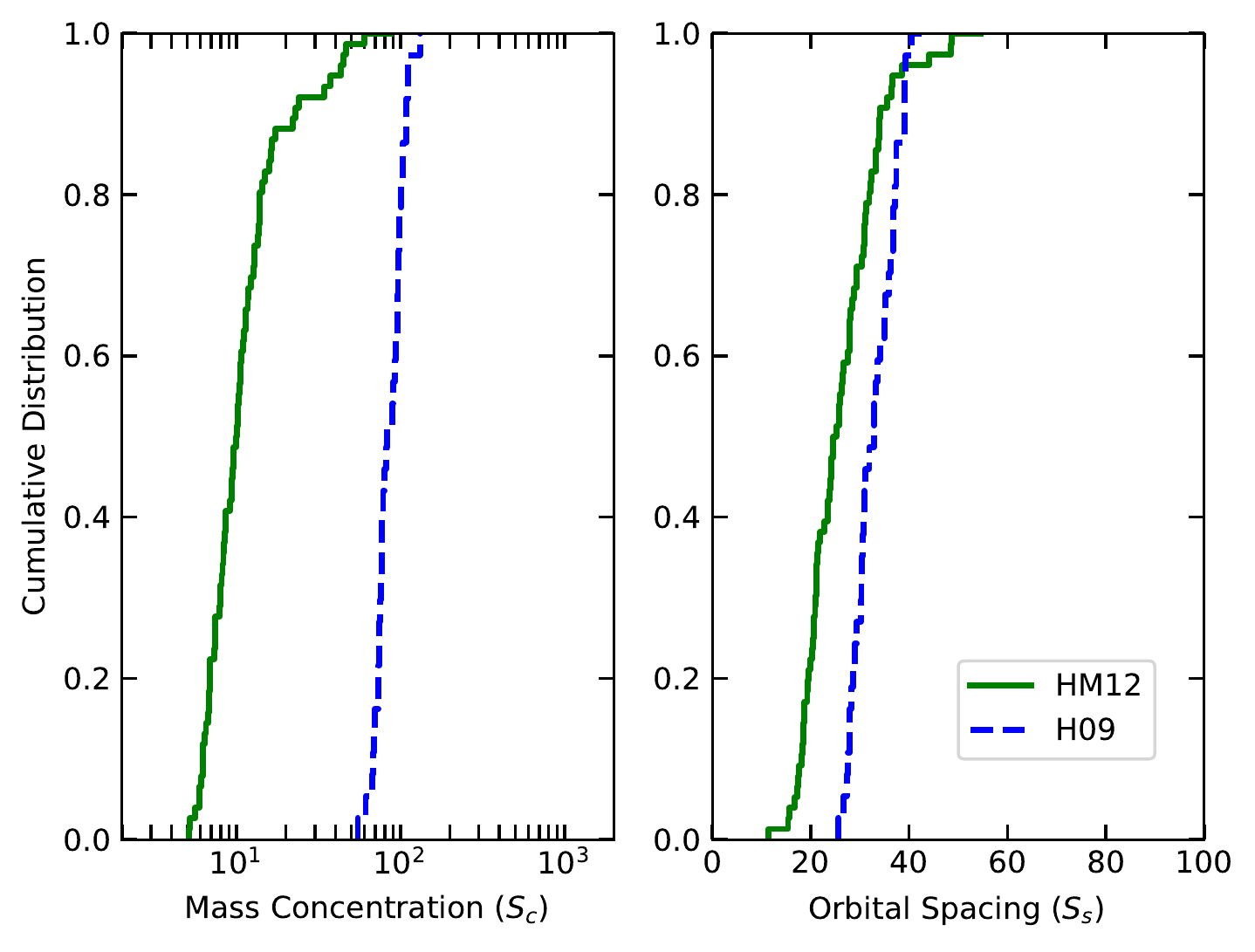}
\caption{The cumulative distributions of $S_c$ and $S_s$ for the simulation results made by H09 and HM12.
The cumulative distribution of $S_c$ shows a clearer difference, 
which comes from different distributions of protoplanets that eventually undergo giant impact.}
\label{fig4}
\end{center}
\end{figure}

\begin{figure}
\begin{center}
\includegraphics[width=8cm]{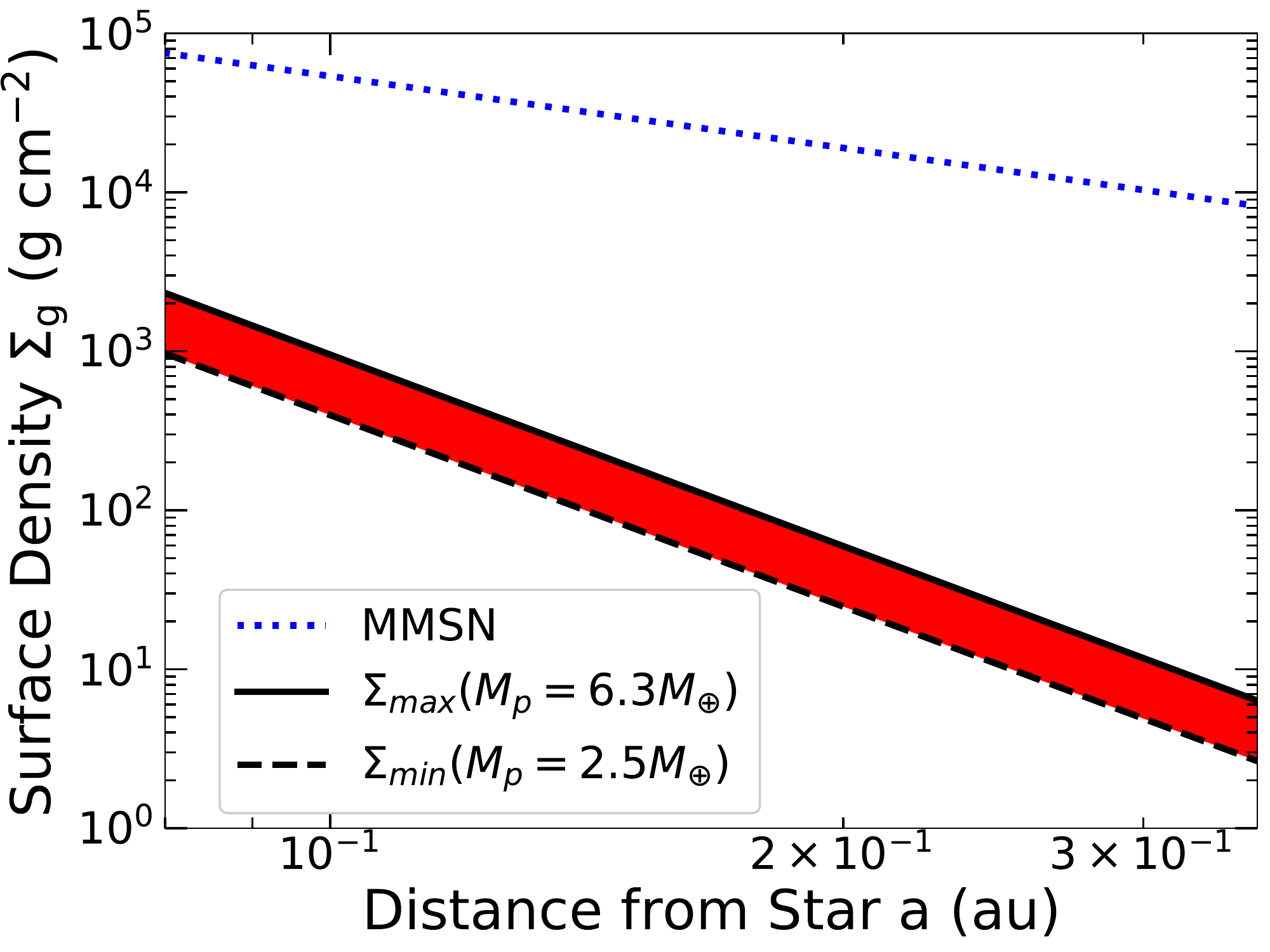}
\caption{The optimal value of $\Sigma_{\rm g}$ as a function of the distance from the central star.
The maximum and minimum values of $\Sigma_{\rm g}$ are computed (see the black solid and dashed lines, respectively),
using the results of Figure \ref{fig3} (see the vertical dashed lines in the left panel).
For comparison, the MMSN model is denoted by the blue dotted line.}
\label{fig5}
\end{center}
\end{figure}

First, we point out that both of our analyses lead to a consistent picture that 
super-Earth (i.e., $M_{p, \rm ave} \la 3 M_{\oplus}$) systems form out of narrow ring-like distributions of protoplanets
while sub-Neptune (i.e., $M_{p, \rm ave} \ga 3 M_{\oplus}$) systems form from radially wide distributions of protoplanets.
The difference in the spatial distribution of protoplanets is produced by planetary migration;
a chain of protoplanets tends to form readily with a wider radial extent for massive protoplanets due to faster migration.
Importantly, the existing $N$-body simulations already show that 
different (narrow vs wide) spatial distributions of protoplanets leads to different (high vs low) values of $S_c$.
Figure \ref{fig4} shows the simulation results obtained by H09 and HM12;
the former targets the formation of terrestrial planets in the solar system and hence starts from narrow rings whose radial extent is comparable to the resulting value of $S_s$.
On the other hand, the latter does the formation of small-sized planets observed by {\it Kepler} and 
begins with wide distributions whose radial extent is a few times larger than the resulting value of $S_s$.

It is obvious that quantitative comparison between the observations (Figure \ref{fig3}) and the simulations (Figure \ref{fig4}) cannot be made 
due to the observational bias and idealization adopted in simulations.
For instance, the observational bias may partially wash out the effect of planetary migration;
this may be a reason of why there is no clear difference in the cumulative distribution of $S_s$ between low-mass and high-mass systems (Figure \ref{fig3}).
However, qualitative comparison may be useful; 
it may be reasonable to consider that the difference in the cumulative distribution of $S_c$ is significant enough to infer the spatial distribution of protoplanets.

Second, we consider the value of $\Sigma_{\rm g}$ in the vicinity of the central star.
As discussed in Section \ref{sec:theory},
a lower value of $\Sigma_{\rm g}$ may be preferred,
in order to limit the effect of planetary migration (Figure \ref{fig2}).
Under the assumption that the population of low-mass and high-mass systems is divided at $M_{p, \rm th}$,
one can compute the optimal value of $\Sigma_{\rm g}$.
Figure \ref{fig5} shows the results.
It is interesting that the resulting values are much lower than the minimum-mass solar nebula (MMSN) model \citep{1981PThPS..70...35H}.
This difference is readily understood by the fact that there is no close-in planet in the solar system,
and hence the MMSN model is not sensitive to such a region.
Furthermore, the recent studies support the lower $\Sigma_{\rm g}$ in the vicinity of the central star \citep[e.g.,][]{2020MNRAS.495.4192C},
which may be caused by stellar magnetic fields \citep{2019A&A...629L...1H} and/or disk winds \citep{2015A&A...584L...1O}.

Finally, we comment on caveats involved in this work.
The most important one would be the observational bias as discussed above.
Since it would be complicated to reliably determine the observational bias for multi-planet systems,
it may be straightforward to run simulations and impose the bias on the simulation results as done by \citet{2020ApJ...897...72M,2020AJ....160..276H}.
This approach allows one to compare simulations and observations directly and quantitatively.
Detailed simulations are desired to verify the results of this work,
which is the target of our future work.

Thus, planetary migration and the properties of protoplanetary disks in the vicinity of the central stars are the key 
to better understanding the origin and properties of observed super-Earths and sub-Neptunes.

\begin{acknowledgments}

The authors thank an anonymous referee for useful comments on our manuscript.
This research was carried out in part at JPL/Caltech, under a contract with NASA.
Y.H. is supported by JPL/Caltech.       

\end{acknowledgments}

\bibliographystyle{aasjournal}
\bibliography{adsbibliography}    



\end{document}